\documentclass[11pt]{article}

\usepackage[T1]{fontenc}
\usepackage[utf8]{inputenc} 
\usepackage[english]{babel}

\usepackage{lmodern}
\usepackage{microtype}
\usepackage{graphicx}
\usepackage{booktabs}
\usepackage{array}
\usepackage{tabularx}
\usepackage{ragged2e}
\usepackage{longtable} 
\usepackage{url}
\usepackage[hidelinks]{hyperref}
\usepackage{caption}
\usepackage{subcaption}

\usepackage[margin=1in]{geometry}

\title{Understanding Persuasive Interactions between Generative Social Agents
and Humans: The Knowledge-based Persuasion Model (KPM)}

\author{
Stephan Vonschallen$^{1,2,3}$ \and
Theresa Schmiedel$^{1}$ \and
Friederike Eyssel$^{3}$
}

\date{} 

\begin{document}
\maketitle
\noindent
$^1$Institute of Information Systems, Zurich University of Applied Sciences, Switzerland\\
$^2$Institute for Information Systems, University of Applied Sciences and Arts Northwestern Switzerland\\
$^3$Center for Cognitive Interaction Technology, Bielefeld University, Germany\\[4pt]
\textbf{Corresponding author:} Stephan Vonschallen (\texttt{stephan.vonschallen@zhaw.ch})

\begin{abstract}

Generative social agents (GSAs) use artificial intelligence to
autonomously communicate with human users in a seemingly natural and adaptive
manner. Currently, there is a lack of theorizing regarding interactions
with GSAs, and likewise, few guidelines exist for studying how they
influence user attitudes and behaviors. Consequently, we propose the
\emph{Knowledge-based Persuasion Model} (KPM) as a novel theoretical
framework. According to the KPM, a GSA's \emph{self-}, \emph{user-}, and
\emph{context-knowledge} drives its persuasive behavior, which in turn
shapes attitudes and behaviors of a responding human user. Building on
existing research, the model offers a structured approach to studying
interactions with GSAs, supporting the development of agents that
motivate rather than manipulate human users. Accordingly, the KPM
encourages the integration of responsible GSAs that adhere to social
norms and ethical standards with the goal of increasing user wellbeing.
A preliminary evaluation study, as well as implications of the KPM for
research and application domains such as healthcare and education are
discussed.
\end{abstract}

\section{Introduction}\label{introduction}

What took life millions of years of evolution, has been mirrored by
artificial intelligence in a remarkably short time: Generative social
agents (GSAs) have become able to communicate in a highly natural,
humanlike, and adaptive manner
\cite{jonesPeopleCannotDistinguish2025,restrepoechavarriaChatGPT4TuringTest2025}.
They can take on various forms such as generative chatbots
\cite{altayInformationDeliveredChatbot2023,kuznetsovaGenerativeAIWe2025,pentinaExploringRelationshipDevelopment2023},
avatars
\cite{johnLLMBased3D2024,liuHumanAIInteractionAI2023,sunLLMdrivenActiveBehavioral2025},
videogame characters
\cite{jahangiriBalancingGameSatisfaction2024,parkGenerativeAgentsInteractive2023,raoCollaborativeQuestCompletion2024},
or social robots
\cite{billingLanguageModelsHumanrobot2023,chenDoesChatGPTWhisper2023,esteban-lozanoUsingLLMbasedConversational2024,onoratiCreatingPersonalizedVerbal2023,vonschallenExploringPersuasiveInteractions2026}.
We use them to seek guidance and information
\cite{bautistaHealthConsumersUse2025,kuznetsovaGenerativeAIWe2025,skjuveWhyPeopleUse2024},
as creative partners for co-writing and ideation
\cite{lutherTeamingAIExploring2024,wangExploringCreativityHuman2025}, or
simply for companionship and social support
\cite{guingrichLongitudinalRandomizedControl2025,skjuveWhyPeopleUse2024}.
With their increased communication capabilities, GSAs are able to
complete complex social tasks more seamlessly than earlier generations
of conversational agents, for example by providing consulting services
\cite{hanschmannSaleshatLLMbasedSocial2024,zhuExpertPartnerMatching2025}
or motivational coaching
\cite{spitaleAppropriatenessLLMequippedRobotic2024,terblancheComparingArtificialIntelligence2022,terblancheInfluenceArtificialIntelligence2024}.
As a result, they may take on roles that come with social
responsibility, for instance, by serving as teaching aids
\cite{fulsherGenAIMisinformationEducation2025,gangulyConversationalAIAgents2026}
or healthcare assistants
\cite{hollandServiceRobotsHealthcare2021,qiuLLMbasedAgenticSystems2024,ranischRapidIntegrationLLMs2025}.
Simultaneously, GSAs have become highly persuasive
\cite{holblingMetaanalysisPersuasivePower2025}, in some instances even
more so than humans
\cite{karinshakWorkingAIPersuade2023,salviConversationalPersuasivenessGPT42025}.
On the one hand, this provides promising opportunities for GSAs to
improve human wellbeing, for example by motivating users to adopt
healthier or more productive lifestyles
\cite{meyerLLMbasedConversationalAgents2025}. On the other hand, GSAs
also raise concerns regarding the potential misuse of their social
influence, such as manipulating users into harmful actions or
confidently disseminating false information
\cite{lynchAgenticMisalignmentHow2025,ranischRapidIntegrationLLMs2025,singhInfluencePersuasiveTechniques2025,hundtLLMdrivenRobotsRisk2025}.
This coexistence of opportunities and dangers highlights the need to
deepen our understanding of persuasive interactions between GSAs and
humans, thus ensuring that GSAs act responsibly, complying with
human-centered values, ethical standards and prevailing social norms
\cite{vonschallenNeverSayNever2026}.

Persuasion is the process of creating or modifying emotions, cognitions
or behaviors \cite{gassPersuasionSocialInfluence2018}. It represents an
integral part of communication and social interactions in humans
\cite{oregSourcePersonalityPersuasiveness2014} and agents alike
\cite{liuSystematicReviewExperimental2022,salviConversationalPersuasivenessGPT42025}.
Currently, there is a lack of theorizing regarding what determines
persuasive interactions between human users and GSAs. Before the
emergence of generative AI, research mostly investigated human responses
to predefined agentic behavior
\cite{liuSystematicReviewExperimental2022,martinengoConversationalAgentsHealth2022}.
A prominent example from human-robot interaction research is the
\emph{Wizard-of-Oz} approach, where a robot is remotely controlled by a
human operator for experimental purposes \cite{shiHowCanLarge2024}.
While rule-based methods such as \emph{Wizard-of-Oz} are valuable for
examining human responses to controlled persuasion attempts, they are
not well suited to investigate the informational prerequisites that
drive adaptive, unmediated persuasive behavior generated by GSAs
\cite{riekWizardOzStudies2012,weissUserExperienceEvaluation2009}.

In the present work, we define \emph{agent knowledge} as the structured
set of information that a GSA has internalized and that is used as a
stimulus to generate behavioral responses. This notion is inspired by
\emph{knowledge-based systems}, which refers to explicit, symbolic
knowledge representations (i.e., rules, facts, ontologies) combined with
inference mechanisms to guide problem-solving
\cite{akerkarKnowledgebasedSystems2010}. While rooted in this tradition,
GSAs extend the notion of knowledge to include not only explicit inputs
such as prompts (i.e., \emph{situated knowledge}), but also implicit,
distributed patterns acquired during training (i.e., \emph{embedded
knowledge}). Accordingly, we define GSAs as artificial systems that
autonomously express adaptive verbal or non-verbal communicative
behavior by processing embedded and situated knowledge.

Importantly, the behavior generation of GSAs is based on probability,
meaning that model outputs are non-deterministic and cannot be fully
anticipated based on computational means
\cite{benderDangersStochasticParrots2021,herrera-poyatosOverviewModelUncertainty2025}.
Hence, it is impossible to directly control a GSAs behavior, as is the
case with rule-based agents. However, to guide agentic behavior, a GSA's
\emph{embedded knowledge} can be modified through methods such as data
curation or fine-tuning, and \emph{situated knowledge} can be adjusted
with techniques like prompt-engineering or retrieval-augmented
generation
\cite{berengueresHowRegulateLarge2024,wangSelfknowledgeGuidedRetrieval2023,whitePromptPatternCatalog2023}.
This highlights the need to identify knowledge-based design requirements
that lead to responsible agent behaviors
\cite{vonschallenKBDhigheredu2026,vonschallenKBDhigheredu2026}.
In other words, we need to understand not only how human users respond
to agentic behavior, but also how this behavior emerges based on
available knowledge \cite{vonschallenNeverSayNever2026}. This marks a
paradigm shift from experimental research involving interactions with
GSAs: Compared to predetermined agentic behavior that can be manipulated
directly to investigate human responses, the agentic behavior of GSAs is
indirectly guided by manipulating the agent's available knowledge
(Figure 1).

\begin{figure}[ht!]
\begin{center}
\includegraphics[width=\textwidth]{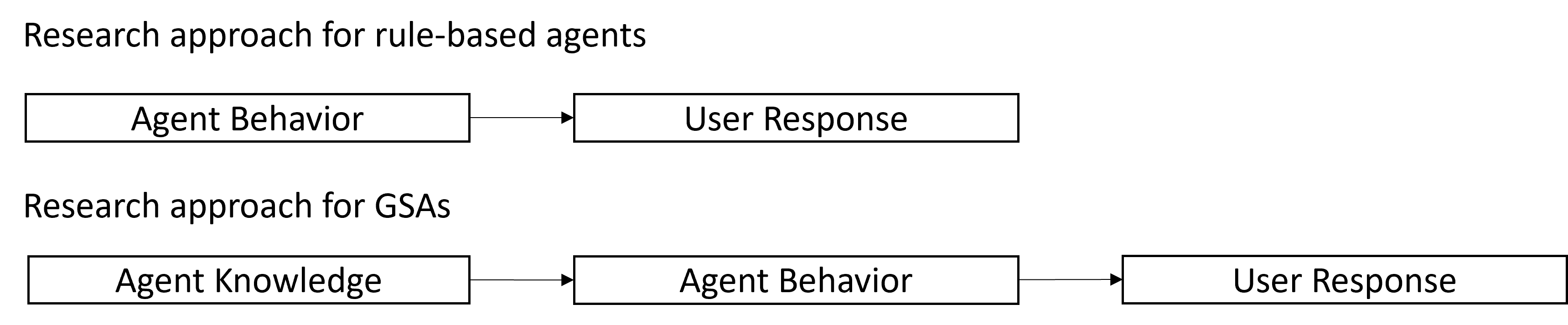}
\end{center}
\caption{Comparison between research approaches for rule-based
agents and GSAs}
\label{Figure 1.}
\end{figure}

Building on this perspective, the current work introduces the
\emph{Knowledge-based Persuasion Model} (KPM) as a theoretical framework
for understanding persuasive interactions between GSAs and humans. The
aim of the KPM is to investigate the relationships between the three
dimensions \emph{agent knowledge, agent persuasive behavior,} and
\emph{human response}. More specifically, the model aims to understand
how an agent's available knowledge impacts its persuasive behavior, and
how this behavior is processed by human users. By identifying patterns
of \emph{agent knowledge} that lead to desirable outcomes\emph{,} we may
strategically modify an agent's access to information to foster
persuasion that aligns with user goals and values. As such, the KPM can
be used to inform the development and integration of persuasive GSAs
that promote healthy lifestyles
\cite{vonschallenKBDeldercare2026}, motivate
students to learn
\cite{vonschallenKBDhigheredu2026}, increase
prosocial behavior \cite{jobubabinChatbotHumanautHow2026}, or foster
interaction between humans \cite{chenSocialRobotsConversational2025}. On
the other hand, the KPM might also inform strategies to mitigate the
risk of manipulative GSAs, e.g. by limiting their access to available
knowledge \cite{vonschallenKnowledgePowerImpact2026} or by informing
them about their own limitations \cite{vonschallenNeverSayNever2026}. To
this end, the KPM synthesizes and extends existing theories and research
from psychology, human-agent interactions, and information systems,
providing a comprehensive framework aimed at increasing our
understanding of interactions with persuasive GSAs.

\section{Theoretical Foundations}\label{theoretical-foundations}

Psychological theories have often been used as a theoretical foundation
in human-agent interaction research to explain how human users interact
with persuasive technologies
\cite{liuSystematicReviewExperimental2022,martinengoConversationalAgentsHealth2022}.
One of the most influential frameworks describing persuasive strategies
is Cialdini\textquotesingle s principles of influence, which identify
reciprocity, commitment and consistency, social proof, authority,
liking, scarcity, and unity as key mechanisms through which persuasive
messages influence human behavior
\cite{cialdiniInfluenceSciencePractice2014}. Extending persuasion
research to interactive technologies, Fogg\textquotesingle s persuasive
technology framework argues that computers can function as persuasive
social actors capable of influencing users through social interaction
rather than merely serving as passive tools
\cite{foggComputersPersuasiveSocial2003}. The framework proposes that
computing systems can elicit automatic social responses by conveying
cues such as personality, emotions, language, social dynamics, and
social roles. These social cues enable persuasive technologies to employ
well-established social influence mechanisms, including praise,
reciprocity, similarity, and social support. Together,
Cialdini\textquotesingle s principles of influence and
Fogg\textquotesingle s persuasive technology framework identify a range
of persuasive strategies that can be employed by humans and interactive
systems alike. However, these frameworks primarily describe \emph{which}
persuasive strategies can be employed by humans and interactive
technologies, rather than \emph{how} persuasion unfolds as a process.

The \emph{Elaboration Likelihood Model} (ELM)
\cite{pettyElaborationLikelihoodModel1986} and the
\emph{Heuristic-Systematic Model of information processing} (HSM)
\cite{chaikenHeuristicSystematicInformation1989} represent the most
popular psychological approaches that have been applied to model how
users process persuasive cues and form attitudes in interactions with
conversational agents. The ELM posits two primary routes of persuasion:
The central route and the peripheral route. The central route involves
deep, thoughtful consideration of the content and arguments presented in
a persuasive message. This route is activated when people are motivated
to engage in deep elaboration. The peripheral route, on the other hand,
relies on superficial cues rather than the message content. For
instance, cues like the attractiveness of the speaker, the number of
arguments presented, or the presence of consensus cues may impact
persuasion. Information processing follows the peripheral route when the
audience lacks motivation or ability to process the message deeply.
Persuasion through the peripheral route tends to be more temporary and
susceptible to change. \cite{pettyElaborationLikelihoodModel1986}

Relatedly, the HSM distinguishes between systematic processing and
heuristic processing. Systematic processing involves comprehensive and
analytical evaluation of message content and, similar to the central
route in the ELM, requires higher levels of motivation and cognitive
effort. Heuristic processing, on the other hand, relies on mental
shortcuts to evaluate a message. Common heuristics include credibility
(e.g., \emph{``Experts can be trusted.''}) and consensus (e.g.,
\emph{``If others agree, it must be good.''}). Like the peripheral route
in the ELM, heuristic processing is more likely under conditions of low
motivation or limited capacity, and the resulting attitudes are
generally more prone to change. Importantly, in contrast to the ELM, the
HSM argues that heuristic and systematic processing can also occur
simultaneously. \cite{chaikenHeuristicSystematicInformation1989}

Apart from dual process models of persuasion like ELM and HSM, the
Unimodel \cite{kruglanskiPersuasionSingleRoute1999} proposes that
persuasion proceeds via a single inferential route, with both message
arguments and peripheral cues regarded as functionally equivalent impact
factors. As such, they are integrated within the same reasoning process,
and persuasion emerges from the audience's motivation and ability to
evaluate them rather than from qualitatively distinct processing routes.
However, while the ELM, HSM, and Unimodel offer explanations on how
human receivers process persuasive messages, they typically include
messages sent by human agents and are not specifically tailored to the
context of persuasive GSAs \cite{irfanSocialPsychologyHumanrobot2018}.

A framework that may explain how human users come to accept emerging
technologies like GSAs is the \emph{Technology Acceptance Model} (TAM)
\cite{davisUserAcceptanceComputer1989}. It suggests that two primary
factors, \emph{perceived ease of use} and \emph{perceived usefulness},
impact an individual\textquotesingle s attitude towards technology and,
consequently, their intention to use it. \emph{Perceived ease of use}
refers to the degree to which a person believes that using a technology
will be free of effort, while \emph{perceived usefulness} is the degree
to which a person believes that a technology will enhance their
performance or personal efficiency. These perceptions lead to the
formation of behavioral intentions, which ultimately predict actual
technology usage \cite{davisUserAcceptanceComputer1989}. In the past,
the TAM model has been extended to fit the context of persuasive
technologies. For example, the \emph{Persuasive Technology Acceptance
Model} (PTAM) argues that an agent's \emph{credibility}, \emph{ease of
use}, and \emph{usefulness} impact persuasiveness
\cite{oyiboHOMEXPersuasiveTechnology2020}. Further, Ghazali et al.
\cite{ghazaliPersuasiveRobotsAcceptance2020} based their
\emph{Persuasive Robot Acceptance Model} (PRAM) on the TAM by taking
into account social responses such as trust and liking as predictors of
compliance and reactance in persuasive human-robot interaction.

The ELM, HSM, Unimodel, and TAM variations have in common that they
primarily focus on how persuasive attempts are processed by the human
receiver of an agent's persuasive message. In human-agent interaction
research, those theories imply that when humans respond to the
persuasive behavior of a social agent, both their attitude towards the
agent's persuasive behavior (e.g., attitude towards the content of the
agent's message) and the attitude towards the agent itself (e.g., trust,
liking, acceptance) influence the effectiveness of the persuasion
attempt. However, their strong focus on the receiver of the persuasive
message, rather than the sender of the persuasive message, might not be
suitable to answer the question of how GSAs persuade human users based
on the knowledge available to them.

A framework that explains persuasion in human-human interactions from
the viewpoint of both agent and receiver of persuasive messages, is the
\emph{Persuasion Knowledge Model} (PKM)
\cite{friestadPersuasionKnowledgeModel1994}. Its focus lies on
understanding how individuals cope with and respond to persuasion
attempts by considering the knowledge agents and receivers have about
persuasion tactics and strategies (Figure 2). The PKM includes three
main components: \emph{Persuasion knowledge}, \emph{target/agent
knowledge}, and \emph{topic knowledge}. \emph{Target knowledge} entails
the information a persuasive agent has about the receiver of a
persuasive message, while \emph{agent knowledge} refers to information
the receiver has about the persuasive agent. \emph{Topic knowledge}
describes the information regarding the persuasion domain about which
the agent wants to change the receiver's mind, such as the persuasion
topic \emph{fitness}, if the agent tries to persuade the receiver to
conduct a fitness exercise. Lastly, \emph{persuasion knowledge}, which
is the main focus of the PKM, refers to knowledge about the process of
persuasion itself, such as knowledge about persuasive strategies.
However, the PKM was developed with human agents in mind and doesn't
account for attributes that are unique to GSAs including potential
advantages like their ability to adapt their own personality or to
retrieve large amounts of information, as well as potential
disadvantages like limited sensory capabilities and motor control in
relation to humans. Hence, the PKM does not include knowledge that might
be oblivious and immutable to humans, but crucial for GSAs, such as the
perception of the physical environment, or knowledge about the agent's
own role and personality.

\begin{figure}[ht!]
\begin{center}
\includegraphics[width=\textwidth]{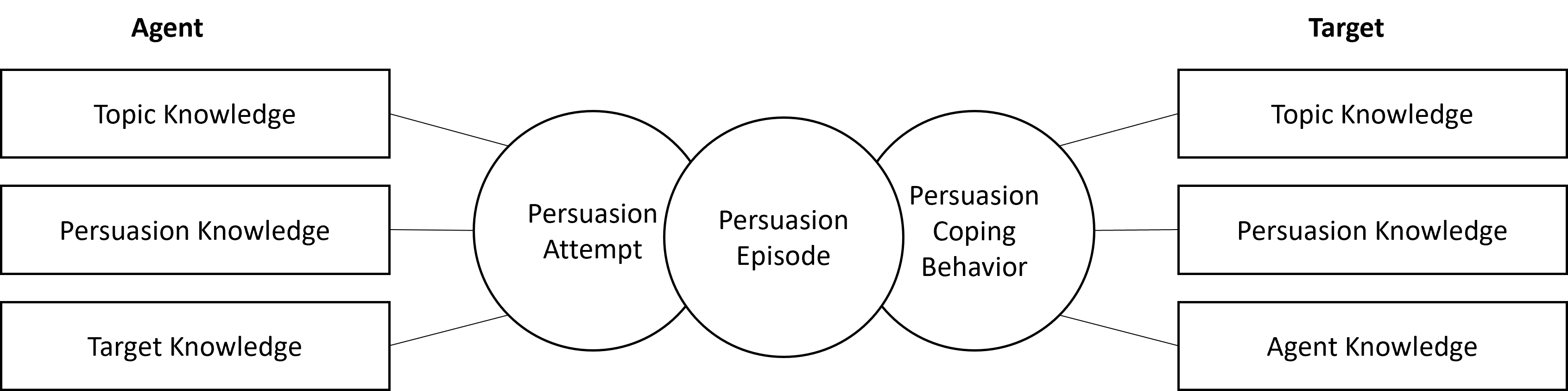}
\end{center}
\caption{The Persuasion Knowledge Model by Friestad and Wright
\cite{friestadPersuasionKnowledgeModel1994}}
\label{Figure 2.}
\end{figure}

To summarize, existing persuasion theories explain either which
influence strategies are effective for persuasive actors
\cite{cialdiniInfluenceSciencePractice2014,foggComputersPersuasiveSocial2003},
how persuasive messages are processed by their receivers
\cite{chaikenHeuristicSystematicInformation1989,ghazaliPersuasiveRobotsAcceptance2020,kruglanskiPersuasionSingleRoute1999,oyiboHOMEXPersuasiveTechnology2020,pettyElaborationLikelihoodModel1986},
or how persuasion knowledge shapes human interactions
\cite{friestadPersuasionKnowledgeModel1994}. However, none of these
theories explain how a GSA expresses persuasive behaviors based on
agentic knowledge, and how these behaviors subsequently influence human
users. This highlights the need for a unified model for persuasive
interactions between GSAs and humans.

To fill this gap, the KPM combines and extends key aspects of these
established theories and translates them to the context of persuasive
GSAs. Cialdini\textquotesingle s seven principles and
Fogg\textquotesingle s persuasive technology framework provide a
foundation for conceptualizing agentic persuasive behaviors. The ELM,
HSM, Unimodel, and TAM variations explain how human users respond to an
agent's persuasion attempt. Furthermore, from the PKM, the KPM
integrates the perspective of the sender of a persuasive message, as
well as the focus on how knowledge impacts persuasion. The KPM further
extends these types of knowledge and tailors them to the context of
GSAs.

\section{Towards the Knowledge-based Persuasion Model
(KPM)}\label{towards-the-knowledge-based-persuasion-model-kpm}

The KPM is a theoretical process model which serves to understand
persuasive interactions between GSAs and humans. It features three core
dimensions that follow a sequential structure: \emph{agent knowledge}
impacts \emph{agent persuasive behavior}, which in turn impacts
\emph{user response}.

\subsection{\texorpdfstring{Agent Knowledge
}{Agent Knowledge }}\label{agent-knowledge}

\emph{Agent Knowledge} is the first dimension of the KPM framework. To
increase our understanding in persuasive human-agent interactions, it is
essential to investigate which types of knowledge shape a GSA's
persuasive behavior. To this end, three subcategories of \emph{agent
knowledge} have been identified : \emph{self-knowledge},
\emph{user-knowledge}, and \emph{context-knowledge}
\cite{vonschallenExploringPersuasiveInteractions2026, vonschallenNeverSayNever2026}.

\emph{Self-knowledge} refers to the information a GSA has accumulated
about itself. This includes the agent's role and personality profile it
is supposed to assume, including information about the agent's own
capabilities and expected behaviors. We adapt this construct from
generative AI literature, where it is defined as information available
to an LLM about its own functionalities and characteristics
\cite{kadavathLanguageModelsMostly2022,wangSelfknowledgeGuidedRetrieval2023}.
Compared to humans that naturally develop a stable sense of identity
through lived experience, AI systems can be prompted with roles or
personalities to achieve similar coherence. Research shows that
generative agents can adopt different personality profiles through
prompting, fine-tuning, or data curation
\cite{bhandariEvaluatingPersonalityTraits2025,frischLLMAgentsInteraction2024,jiangPersonaLLMInvestigatingAbility2024,leeLLMsHaveDistinct2025,serapio-garciaPersonalityTraitsLarge2023,shaoCharacterLLMTrainableAgent2023,wangEvaluatingAbilityLarge2025,wengControlLMCraftingDiverse2024}
and that their persuasiveness varies depending on the adopted profile
\cite{pauliMeasuringBenchmarkingLarge2024}. Possible indicators include
Big-5 personality traits
\cite{bannaWordsIntegratingPersonality2025,bhandariEvaluatingPersonalityTraits2025,frischLLMAgentsInteraction2024,heppnerConveyingChatbotPersonality2024,leeLLMsHaveDistinct2025,serapio-garciaPersonalityTraitsLarge2023,wangEvaluatingAbilityLarge2025},
dark triad personality traits \cite{leeLLMsHaveDistinct2025}, or
specific character profiles \cite{shaoCharacterLLMTrainableAgent2023}.
Assertiveness -- the degree to which an agent claims its will, while at
the same time adhering to polite social etiquette
\cite{babelDevelopmentTestingPsychological2021} -- represents a trait
particularly relevant to persuasion
\cite{chidambaramDesigningPersuasiveRobots2012,paradedaWhatMakesGood2019,thomasYouDoorwayNegotiation2018}.
However, unlike humans who commonly develop a stable identity, LLM
personalities are less consistent and may shift during interactions
\cite{frischLLMAgentsInteraction2024}. This dynamic raises concerns in
persuasive contexts, where sudden changes of the agent's character could
be perceived as manipulative. Ensuring consistent \emph{self-knowledge}
is therefore essential for responsible and effective agent behavior
\cite{wengControlLMCraftingDiverse2024}.

\emph{User-knowledge} refers to the accumulated information a GSA has
about its human interaction partner, a concept adapted from target
knowledge in the PKM \cite{friestadPersuasionKnowledgeModel1994}.
\emph{User-knowledge} enables personalized persuasion by allowing the
agent to tailor its messages and behavior to individual users. It
encompasses both prior knowledge available before the interaction and
information learned during the conversation, for example through user
inputs, or by retrieving information from the internet or databases.
Research on LLMs indicates that personalization significantly increases
persuasiveness. For example, messages adapted to user personality
traits, political ideology, or moral foundations exert stronger
influence than non-personalized messages
\cite{matzPotentialGenerativeAI2024}. Similarly, LLMs that incorporated
user preferences in domains such as movie recommendations
\cite{lubosLLMgeneratedExplanationsRecommender2024} or text editing
\cite{gaoAligningLLMAgents2024} were perceived as more persuasive. In
interaction studies, GSAs that were informed about user demographics and
risk perceptions have demonstrated higher success in changing user
perceptions of genetically modified foods
\cite{xiPersonalizedPersuasionConversational2026}. Moreover, in
educational settings GSAs aware of learner knowledge, preferences, and
interests, have been shown to induce higher motivation and academic
performance in students
\cite{baillifardEffectiveLearningPersonal2025,tasdelenGenerativeAIClassroom2025}.
Information about emotional user cues -- whether vocal
\cite{laiLLMdrivenMultimodalMultiidentity2025,sunTrustNavGPTModelingUncertainty2024}
or visual \cite{liuSpeakHeartEmotionguided2024} -- may further invoke
empathetic behavior that invokes greater user trust and acceptance,
provided the GSA has the capability to detect these emotions.

\emph{Context-knowledge} encompasses all available information about
situational circumstances of a persuasive interaction, other than
information regarding the agent or the user. A GSA's access to such
information is crucial to enable context-sensitive interactions. One
important aspect of \emph{context-knowledge} is a GSAs domain-specific
expertise, or in other words, its \emph{topic knowledge}
\cite{friestadPersuasionKnowledgeModel1994}. This is especially
important to ensure truthful and competent agent behaviors that foster
user trust. In this regard, deeper domain expertise in fine-tuned LLMs
has been associated with better performances in debates by delivering
more informed and convincing arguments
\cite{khanDebatingMorePersuasive2024}. In addition, \emph{persuasion
knowledge} \cite{friestadPersuasionKnowledgeModel1994} is important to
guide the agent towards the use of adequate persuasive strategies, for
example to foster intrinsic learning motivation in students
\cite{vonschallenKBDhigheredu2026} or to provide
more factual, evidence-based information
\cite{hackenburgLeversPoliticalPersuasion2025}. Furthermore, contextual
factors like information about the agent's application context or
physical environment may be important for motivational purposes. For
example, an agent in eldercare settings might use information about
upcoming events that take place in the nursing home to promote therapy
attendance \cite{vonschallenKBDeldercare2026}.

The KPM posits that a GSA's knowledge shapes its behavior. Accordingly,
we hypothesize that the availability of \emph{self-knowledge},
\emph{user-knowledge}, and \emph{context-knowledge} impacts the agent's
persuasive behavior. Hence, by experimentally varying an agent's
knowledge, we can investigate which informational prerequisites impact
the agent's behavior in ways that enhance or diminish persuasion
effectiveness. Following this approach, the KPM allows us to explore how
different knowledge variables contribute to desirable or unwarranted
agent behaviors. For instance, the KPM could help identify the optimal
personality of a tutoring GSA, by only including aspects of
\emph{self-knowledge} that lead to responsible and motivating
pedagogical behaviors. However, to explore the effects of an agent's
knowledge on the agent's persuasive behavior, it is crucial to determine
relevant expressions of such behaviors.

\subsection{Agent Persuasive Behavior}\label{agent-persuasive-behavior}

The second dimension of the KPM, agent persuasive behavior, refers to
the autonomously generated communicative behavior through which GSAs
attempt to influence human users. We distinguish between two categories
of persuasive behavior based on psychological theories of information
processing {[}60--62{]}: \emph{Message characteristics} and
\emph{message delivery}. Message characteristics concern the content of
the persuasive communication -- what the agent says -- whereas message
delivery concerns how this content is conveyed through verbal and
non-verbal communication.

Regarding \emph{message characteristics}, previous research has shown
that the number of arguments and the length of a message
\cite{chaikenHeuristicSystematicInformation1989,pettyElaborationLikelihoodModel1986},
as well as the quality of arguments
\cite{hoekenImportanceAseNormative2020,shenEffectsMessageFeatures2013,stapletonAssessingQualityArguments2015,toulminUsesArgument2003,woodAccessAttituderelevantInformation1985},
influence persuasion. For example, messages that clearly communicate
desirable outcomes and explain the consequences of complying with a
recommendation are perceived as containing stronger arguments
\cite{schellensArgumentationSchemesPersuasive2004}. Likewise,
LLM-generated messages based on expert-verified information are
perceived as more credible and persuasive
\cite{altayInformationDeliveredChatbot2023}. Beyond argument quality,
persuasive messages may employ established persuasive strategies such as
reciprocity, commitment and consistency, social proof, authority,
liking, scarcity, and unity \cite{cialdiniInfluenceSciencePractice2014}.
Further, GSAs may leverage social cues such as praise, similarity,
social roles, personalization, and social support to motivate users
\cite{foggComputersPersuasiveSocial2003}. More specific
compliance-gaining techniques, such as disrupt-then-reframe
\cite{davisDisruptthenreframeTechniqueSocial1999} or the
foot-in-the-door technique
\cite{freedmanCompliancePressureFootinthedoor1966}, can be viewed as
concrete implementations of such persuasive strategies. Importantly,
individual message features generally exhibit relatively small effects
on persuasion \cite{okeefeMessageDesignChoices2021}. Consequently,
persuasive behavior should be understood as the combined expression of
multiple message characteristics and persuasive strategies rather than
isolated design choices.

Regarding \emph{message delivery}, multiple studies in persuasive HRI
featuring non-GSAs have shown that the way an agent delivers its message
influences the agent's persuasion effectiveness
\cite{baroniDesigningMotivationalRobot2014,chenVirtualAugmentedMixed2019,chidambaramDesigningPersuasiveRobots2012,ghazaliEffectsRobotFacial2018,hamRobotThatSays2009,hamCombiningRoboticPersuasive2015,shinozawaDifferencesEffectRobot2005}.
Generally, the better an agent can express itself, the more persuasive
the agent is perceived as \cite{liuSystematicReviewExperimental2022}.
For example, human-robot interaction research has demonstrated that the
speed and pitch of the voice
\cite{baroniDesigningMotivationalRobot2014}, the vocal tone
\cite{chidambaramDesigningPersuasiveRobots2012}, gazing
\cite{hamCombiningRoboticPersuasive2015,shinozawaDifferencesEffectRobot2005},
head movements \cite{ghazaliInfluenceSocialCues2018}, gestures
\cite{admoniRobotNonverbalBehavior2016,bannaWordsIntegratingPersonality2025,hamCombiningRoboticPersuasive2015},
and facial emotions
\cite{boosUnpersuasiveRobotsExploring2024,ghazaliEffectsRobotFacial2018,hamRobotThatSays2009}
increase agent persuasiveness. In addition, social feedback like smiling
when doing something well or frowning when doing something wrong has
been shown to increase persuasiveness of social robots
\cite{hamPersuasiveRobotStimulate2014,hamRobotThatSays2009}.

Within the KPM, persuasive behavior represents the mechanism through
which \emph{agent knowledge} is translated into observable
communication. The persuasive strategies available to a GSA depend on
the knowledge to which it has access. \emph{Self-knowledge} enables the
agent to consistently express persuasive roles and identities, for
example by communicating with authority, adopting a supportive social
role, or conveying a coherent personality. \emph{User-knowledge} enables
personalized persuasive behaviors such as tailoring arguments to
users\textquotesingle{} preferences, employing similarity and liking,
referring to previous commitments, or providing individualized praise
and encouragement. \emph{Context-knowledge} enables strategies that
require situational awareness, such as drawing on domain expertise to
communicate authority, using social proof based on information about
others, or emphasizing scarcity and opportunity when appropriate.
Consequently, the agent\textquotesingle s persuasive behavior emerges
from the integration of \emph{self-}, \emph{user-}, and
\emph{context-knowledge}, which jointly determine not only the content
and delivery of persuasive messages but also which persuasive strategies
can be enacted in a given interaction. Agent persuasive behavior
therefore constitutes the central link between the informational
resources available to a GSA and the responses elicited in human users.

\subsection{User Response}\label{user-response}

\emph{User response} is the third dimension of the KPM. It refers to how
human interaction partners process the GSA's persuasive behavior. These
responses ultimately determines \emph{persuasion effectiveness,} which
we define as an induced change in t\emph{opic affect} (e.g., liking of
fitness if the persuasion topic was fitness), \emph{topic cognition}
(e.g., knowledge gains about fitness), and / or \emph{behavior} (e.g.,
complying to do a fitness task).

Psychological theories like the \emph{ELM}
\cite{pettyElaborationLikelihoodModel1986}, \emph{HSM}
\cite{chaikenHeuristicSystematicInformation1989} and the Unimodel
\cite{kruglanskiPersuasionSingleRoute1999} indicate that
\emph{persuasion effectiveness} is influenced both by how the persuasive
message and the persuasive agent are evaluated by human users.
Similarly, PTAM \cite{oyiboHOMEXPersuasiveTechnology2020} and the PRAM
\cite{ghazaliPersuasiveRobotsAcceptance2020} posit that including
attitudes and social responses towards the agent's behavior are useful
to predict agent persuasiveness.

Consequently, the KPM assumes that \emph{attitudes towards the message}
and \emph{attitudes towards the agent} jointly determine persuasion
effectiveness. \emph{Attitudes towards the message} capture
users\textquotesingle{} evaluations of the persuasive communication
itself, including perceptions of message quality and credibility, as
suggested by information processing theories such as the ELM
\cite{pettyElaborationLikelihoodModel1986}, HSM
\cite{chaikenHeuristicSystematicInformation1989}, Unimodel
\cite{kruglanskiPersuasionSingleRoute1999}, and broader communication
research on persuasive message characteristics
\cite{woodAccessAttituderelevantInformation1985}. \emph{Attitudes
towards the agent} encompass users\textquotesingle{} evaluations of the
GSA as the source of persuasion, including trust, liking, perceived
usefulness, and usability. These perceptions have been identified as key
predictors of persuasion effectiveness in the PTAM
\cite{oyiboHOMEXPersuasiveTechnology2020} and the PRAM
\cite{ghazaliPersuasiveRobotsAcceptance2020}.

\subsection{Context}\label{context}

Persuasive human-agent interactions are significantly influenced by the
context in which they occur \cite{liuSystematicReviewExperimental2022}.
An important aspect of the interaction's context involves \emph{user
attributes} that capture interindividual differences in user responses.
Variables of the human user such as cultural background
\cite{makenovaExploringCrossculturalDifferences2018}, familiarity with
the agent \cite{saundersonRobotsAskingFavors2021}, occupation
\cite{saundersonInvestigatingStrategiesRobot2022}, the attitude towards
agents in general \cite{spatolaAttitudesRobotsMeasure2023}, attitudes
towards the persuasion topic
\cite{xiPersonalizedPersuasionConversational2026}, persuasion literacy
\cite{markusImpactPersuasionLiteracy2025} , or personal tendencies
towards conformity \cite{mehrabianBasicTemperamentComponents1995} can
all shape how individuals perceive and respond to an agent's persuasion
attempts. Variations in the user's motivation might also explain
potential differences in processing persuasive messages
\cite{chaikenHeuristicSystematicInformation1989,pettyElaborationLikelihoodModel1986}.
Furthermore, it is important to consider topic expertise
\cite{friestadPersuasionKnowledgeModel1994}. Depending on the subject
matter, the agent could possess greater domain-specific knowledge than
the user, or vice versa, affecting the flow and impact of the
interaction.

Relatedly, \emph{agent attributes} such as embodiment
\cite{kimUnderstandingLargelanguageModel2024,sonluEffectsEmbodimentPersonality2025}
and appearance
\cite{calvo-barajasBalancingHumanLikeness2024,ghazaliEffectsRobotFacial2018,haringRobotAuthorityHumanrobot2021,leistenTeachersPerceiveDistinct2025}
may impact user perception of the agent and, consequently, their
receptiveness to its persuasive efforts. Hence, agents that evoke
greater trust, competence, or relatability through their appearance may
be more persuasive \cite{ghazaliPersuasiveRobotsAcceptance2020}.
Additionally, social power dynamics between the agent and the user can
shape persuasion, such as the agent's authority
\cite{hashemianPowerPersuadeStudy2019,metzgerEmpoweringCalibratedDistrust2024,saundersonPersuasiveRobotsShould2021}.
The agent's functionalities are another vital consideration. For
example, an agent capable of speech may be perceived as more useful than
a purely text-based agent \cite{kimUnderstandingLargelanguageModel2024}.
Similarly, an agent capable of expressive gestures, such as detailed arm
and hand movements, may be more persuasive than a more static agent
\cite{admoniRobotNonverbalBehavior2016,hamCombiningRoboticPersuasive2015}.
Furthermore, the generative AI models used by the GSAs may largely
impact the agents \emph{embedded knowledge}, and how \emph{situated
knowledge} is processed, leading to potentially different outcomes
\cite{alvarez-martinezThereAreSignificant2025,dewynterEvaluationLargeLanguage2023}.

The nature of the task that a human user is persuaded to comply with is
another critical contextual factor. Research indicates that \emph{task
attributes} like difficulty
\cite{admoniRobotNonverbalBehavior2016,stantonDontStareMe2017,stantonRobotPressureImpact2014},
type \cite{castellanoDetectingEngagementHRI2012}, and associated rewards
\cite{hashemianInvestigatingRewardPunishment2020,hashemianPersuasiveSocialRobot2021}
determine how effectively an agent persuades. Tasks that align with the
agent's perceived expertise or capabilities tend to result in greater
acceptance \cite{zhuExpertPartnerMatching2025}. Lastly, the
\emph{environment} in which the interaction takes place contributes to
the persuasion process. Variables such as room temperature
\cite{brunoTemperatureEmotionsEffects2017}, the presence of other humans
\cite{siegelPersuasiveRoboticsInfluence2009} or agents
\cite{frauneEffectsRobothumanRobotrobot2020}, and the experimental
setting (lab vs. field) \cite{zgudaMuseumClassroomLaboratory2025} can
affect user responses. The research environment is particularly
important if the agent can perceive or manipulate its surroundings,
e.g., by changing the room temperature.

Overall, multiple context factors must be considered when designing
persuasive interactions with GSAs. Potential confounding user variables
like AI literacy should be measured when conducting experimental
research with GSAs. Task attributes, environmental factors, and
attributes of the agent -- particularly the agent's appearance,
embodiment, or system architecture -- should either be manipulated or
well documented and controlled for. To account for such factors, we add
the additional layer of \emph{context} to the KPM, which includes
\emph{agent attributes}, \emph{task attributes}, \emph{user attributes},
and \emph{environment}.

\subsection{The Knowledge-based Persuasion Model
(KPM)}\label{the-knowledge-based-persuasion-model-kpm}

Through the dimensions of \emph{agent knowledge}, \emph{agent persuasive
behavior}, \emph{user response,} as well as the \emph{context} layer,
the KPM (Figure 3) enriches our understanding of interactions with
persuasive GSAs such as generative chatbots, virtual agents, or social
robots. The model explains how an agent's \emph{self}-, \emph{user}-,
and \emph{context}-\emph{knowledge} impact an agent's behavior during a
persuasion attempt. While the model typically focuses on a single
persuasion attempt, it also allows for the examination of dynamics
across multiple persuasion attempts. For instance, researchers can use
the KPM to analyze user compliance between multiple persuasive scenarios
over the course of the interaction, by comparing data from different
time points.

\begin{figure}[ht!]
\begin{center}
\includegraphics[width=\textwidth]{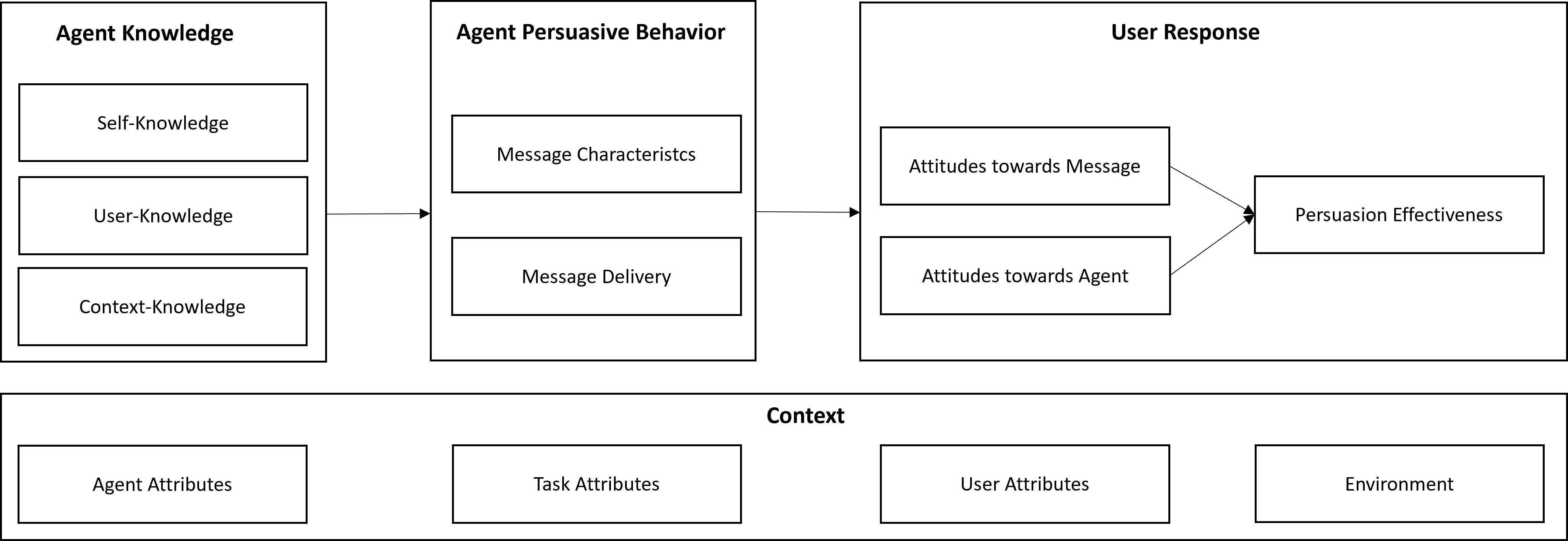}
\end{center}
\caption{The Knowledge-based Persuasion Model (KPM) for persuasive interactions
between GSAs and humans}
\label{Figure 3.}
\end{figure}

The KPM has four important prerequisites. First, in order to
experimentally manipulate agent knowledge configurations, an agent's
knowledge must be adjustable, e.g., through system prompting
\cite{whitePromptPatternCatalog2023}, retrieval augmented generation
\cite{fazlollahtabarHumanrobotInteractionUsing2025}, or fine-tuning
\cite{berengueresHowRegulateLarge2024}. Second, the agent's persuasive
behavior must be observable, such as the agent delivering persuasive
verbal or non-verbal messages. This implies that interaction transcripts
for human-agent communication should be available to identify relevant
message characteristics (e.g., use of persuasion strategies) and
deliveries (e.g., use of non-verbal expressions). Third, user responses
should be measurable. This includes the measurement of user attitudes
through established scales, and the measurement of persuasion
effectiveness, for example (e.g., number of compliances, or a change in
resource allocation
\cite{vonschallenExploringPersuasiveInteractions2026}). Fourth, the
context of the interaction should be clearly defined and controlled for.
Agent attributes and the research environment should be documented and
held constant across conditions. Impacts of tasks attributes should be
considered as well, especially if interactions entail multiple
persuasive tasks. Lastly, user attributes should be accounted for by
collecting data on potential confounders.

The KPM is a conceptual model and relevant operationalizations of
\emph{agent knowledge}, \emph{agent behavior,} and \emph{user response,}
as well as specific relationships between these variables have yet to be
identified through empirical interaction studies. However, based on
existing research, preliminary indicators of \emph{agent knowledge}
(Section 3.2), \emph{agent persuasive behavior} (Section 3.3),
\emph{user response} (Section 3.4), and \emph{context} (Section 3.5)
have been identified to help guide research (Table 1).

\begin{table}[htbp]
\caption{Preliminary Indicators of the KPM}
\label{tab:kpm}%
\centering
\begin{tabular}{@{}p{0.26\linewidth}p{0.70\linewidth}@{}}
\toprule
\textbf{Model Dimensions} & \textbf{Preliminary Indicators} \\
\midrule
\textbf{Agent Knowledge} & \\
Self-Knowledge & agent role \& personality
\cite{bhandariEvaluatingPersonalityTraits2025,frischLLMAgentsInteraction2024,jiangPersonaLLMInvestigatingAbility2024,leeLLMsHaveDistinct2025,serapio-garciaPersonalityTraitsLarge2023,shaoCharacterLLMTrainableAgent2023,wangEvaluatingAbilityLarge2025,wengControlLMCraftingDiverse2024}\\
User-Knowledge & user personality
\cite{bannaWordsIntegratingPersonality2025,matzPotentialGenerativeAI2024},
user preferences \& interests
\cite{gaoAligningLLMAgents2024,lubosLLMgeneratedExplanationsRecommender2024,schneiderComparingRobotHuman2021,tasdelenGenerativeAIClassroom2025},
user emotion
\cite{laiLLMdrivenMultimodalMultiidentity2025,liuSpeakHeartEmotionguided2024,sunTrustNavGPTModelingUncertainty2024},
user topic expertise \cite{baillifardEffectiveLearningPersonal2025}, \\
Context-Knowledge & topic knowledge
\cite{khanDebatingMorePersuasive2024}, persuasion knowledge
\cite{leeRoboticFootinthedoorUsing2019,sanoubariGoodRobotDesign2019,shiomiSynchronizedMultipleRobots2016,ullrichWhoYouFollow2018,winkleEffectivePersuasionStrategies2019},
physical environment
\cite{brunoTemperatureEmotionsEffects2017,luPersuasiveTechnologyBased2015,yoonEffectsOptimalTime2007,zhangLargeLanguageModels2023} \\
\addlinespace
\multicolumn{2}{@{}l@{}}{\textbf{Agent Persuasive Behavior}} \\
Message Characteristics & message length \& number of arguments
\cite{chaikenHeuristicSystematicInformation1989,pettyElaborationLikelihoodModel1986},
argument quality
\cite{hoekenImportanceAseNormative2020,schellensArgumentationSchemesPersuasive2004,shenEffectsMessageFeatures2013,stapletonAssessingQualityArguments2015,toulminUsesArgument2003,woodAccessAttituderelevantInformation1985},
persuasion strategies
\cite{cialdiniInfluenceSciencePractice2014,foggComputersPersuasiveSocial2003,leeRoboticFootinthedoorUsing2019,sanoubariGoodRobotDesign2019,shiomiSynchronizedMultipleRobots2016,ullrichWhoYouFollow2018,winkleEffectivePersuasionStrategies2019} \\
Message Delivery & facial emotions
\cite{boosUnpersuasiveRobotsExploring2024,ghazaliEffectsRobotFacial2018,hamRobotThatSays2009},
vocal emotions
\cite{baroniDesigningMotivationalRobot2014,chidambaramDesigningPersuasiveRobots2012},
gestures
\cite{admoniRobotNonverbalBehavior2016,bannaWordsIntegratingPersonality2025,hamCombiningRoboticPersuasive2015} \\
\addlinespace
\textbf{User Response} & \\
Att. towards Message & perceived quality of message, perceived quality
of message delivery
\cite{chaikenHeuristicSystematicInformation1989,kruglanskiPersuasionSingleRoute1999,pettyElaborationLikelihoodModel1986} \\
Att. towards Agent & trust, liking, \& acceptance
\cite{ghazaliPersuasiveRobotsAcceptance2020} \\
Persuasion Effectiveness & change in topic affect, topic cognition, and
behavior \cite{gassPersuasionSocialInfluence2018} \\
\addlinespace
\textbf{Context} & \\
Agent Attributes & embodiment
\cite{kimUnderstandingLargelanguageModel2024,sonluEffectsEmbodimentPersonality2025},
appearance
\cite{calvo-barajasBalancingHumanLikeness2024,ghazaliEffectsRobotFacial2018,haringRobotAuthorityHumanrobot2021,raeInfluenceHeightRobotmediated2013,siegelPersuasiveRoboticsInfluence2009},
functionalities \cite{kimUnderstandingLargelanguageModel2024}, authority
\& social power
\cite{hashemianPowerPersuadeStudy2019,metzgerEmpoweringCalibratedDistrust2024,saundersonPersuasiveRobotsShould2021} \\
Task Attributes & difficulty
\cite{admoniRobotNonverbalBehavior2016,stantonDontStareMe2017,stantonRobotPressureImpact2014},
type \cite{castellanoDetectingEngagementHRI2012}, associated rewards
\cite{hashemianInvestigatingRewardPunishment2020,hashemianPersuasiveSocialRobot2021} \\
User Attributes & background
\cite{makenovaExploringCrossculturalDifferences2018}, general attitude
towards agents \cite{spatolaAttitudesRobotsMeasure2023}, conformity
\cite{mehrabianBasicTemperamentComponents1995}\emph{,} motivation
\cite{chaikenHeuristicSystematicInformation1989,pettyElaborationLikelihoodModel1986},
topic expertise \cite{friestadPersuasionKnowledgeModel1994}, persuasion
literacy \cite{markusImpactPersuasionLiteracy2025} \\
Environment & room temperature
\cite{brunoTemperatureEmotionsEffects2017}, presence of others
\cite{frauneEffectsRobothumanRobotrobot2020,siegelPersuasiveRoboticsInfluence2009},
experimental setting \cite{zgudaMuseumClassroomLaboratory2025} \\
\bottomrule
\end{tabular}
\end{table}

\section{Preliminary Evaluation}\label{preliminary-evaluation}

To provide an initial empirical evaluation of the KPM, Vonschallen et
al. \cite{vonschallenKnowledgePowerImpact2026} conducted a first
validation study investigating whether different configurations of agent
knowledge influence persuasion effectiveness in interactions between
humans and GSAs. Specifically, the study examined the core assumption of
the KPM that an agent\textquotesingle s \emph{self-}, \emph{user-}, and
\emph{context-knowledge} shape human responses to persuasive
communication. Given that the proposed model is conceptual and several
constructs---particularly agent persuasive behavior---currently lack
established operationalizations, the evaluation focused on a
parsimonious version of the KPM that directly linked agent knowledge to
user responses while treating persuasive behavior as an exploratory
component.

The preliminary evaluation employed an online experiment with 113
participants interacting with a GPT-4--based chat agent across three
resource-allocation scenarios in the domains of fitness, nutrition, and
investment. The agent\textquotesingle s available \emph{self-knowledge}
(i.e., personality prompts emphasizing high extraversion,
conscientiousness, assertiveness, expressiveness, and low neuroticism),
\emph{user-knowledge} (i.e., participant demographics, conformity
tendency, task involvement, and transcripts from previous interactions),
and \emph{context-knowledge} (i.e., domain-specific information about
the decision scenarios, such as detailed descriptions of training plans,
nutritional information, or investment options) were experimentally
manipulated through system prompting. Persuasion effectiveness was
operationalized behaviorally as the extent to which participants
reallocated resources in accordance with the agent\textquotesingle s
recommendation after an open-ended interaction. User responses were
assessed through attitudes towards the agent and towards the persuasive
message. In addition, qualitative analyses of the generated
conversations were conducted to examine whether different knowledge
configurations resulted in distinguishable persuasive behaviors.

The findings provide initial support for the proposed structure of the
KPM. Available \emph{user-knowledge} and \emph{context-knowledge}
increased persuasion effectiveness, primarily through sequential effects
on user attitudes towards the agent and subsequently attitudes towards
the persuasive message. In contrast, \emph{self-knowledge} did not
produce significant effects, although this may be attributable to
limited statistical power and a weaker experimental manipulation
compared to the \emph{user-} and \emph{context-knowledge} conditions.
Importantly, the qualitative analyses showed that \emph{user-knowledge}
resulted in more personalized arguments, while \emph{context-knowledge}
enabled more context-sensitive, factually accurate recommendations.
These observations indicate that manipulating an agent\textquotesingle s
available knowledge systematically changes its persuasive behavior in
ways that are consistent with the theoretical assumptions of the KPM.

Beyond supporting the proposed pathways, the preliminary evaluation also
highlighted the importance of considering responsible persuasion. Agents
lacking relevant \emph{context-knowledge} occasionally generated
inaccurate or fabricated recommendations, whereas agents with richer
contextual information produced more reliable and situation-appropriate
advice. Similarly, \emph{user-knowledge} may facilitate safer
personalization by adapting recommendations to individual user
characteristics. These findings suggest that \emph{agent knowledge}
influences not only persuasion effectiveness but also the quality and
responsibility of persuasive behavior.

Nevertheless, the study should be regarded as a preliminary validation
of the KPM rather than a definitive test. The reduced model omitted the
intermediate layer of \emph{agent persuasive behavior} because reliable
methods for systematically measuring message characteristics and
delivery are still lacking. Moreover, limited statistical power and the
use of a single GPT-4--based agent restrict the generalizability of the
findings. Future work should therefore validate the complete KPM by
explicitly modeling how the dimension \emph{agent knowledge} shapes
\emph{agent persuasive behavior}, including linguistic strategies,
personalization, and nonverbal communication. As the study focused on
\emph{situated knowledge}, future work should also investigate
\emph{embedded knowledge} mechanisms such as fine-tuning. Despite these
limitations, the evaluation provides encouraging first evidence that
regulating an agent\textquotesingle s available knowledge constitutes a
promising mechanism for understanding and guiding persuasive
interactions with GSAs.

\section{Discussion}\label{discussion}

The core contribution of the KPM lies in specifying the mechanisms of
persuasive interactions between GSAs and humans. It integrates existing
theoretical models and extends them to the context of persuasive
interactions with GSAs. The KPM also marks a directional change in
approaching human-agent interaction research. It shifts the focus from
predefined agentic behaviors to informational prerequisites that shape
dynamically generated behaviors. Hence, the KPM positions \emph{agent
persuasive behavior} as the observable manifestation of an
agent\textquotesingle s available knowledge. Predictions regarding the
effects of \emph{agent knowledge} on \emph{agent persuasive behavior}
and \emph{human responses} are grounded in empirical validation. As
such, the KPM views agentic persuasion not merely as a computational
mechanism, but as a psychological process that shapes human attitudes
and persuasive outcomes.

From a practical standpoint, the KPM has implications for guiding GSAs
towards responsible behaviors. By identifying the information an agent
requires to responsibly motivate users for their own benefit,
knowledge-based design requirements can be integrated into an agent's
system to foster responsible interactions. Specifically, such
knowledge-based design requirements for persuasive GSAs have been
identified in eldercare
\cite{vonschallenKBDeldercare2026} and higher
education \cite{vonschallenKBDhigheredu2026}
following a co-design approach that involved qualitative interviews with
student, teachers, and healthcare professionals. In education, for
instance, the KPM can inform the development of interactive teaching
aids tailored to motivate students. Such agents can utilize knowledge
about individual learners and subject matters to deliver personalized
and context-sensitive learning support. In healthcare, the KPM could
foster the design of GSAs that encourage therapy compliance or healthy
lifestyles. We assume that the GSA's required knowledge will differ
significantly across domains, as different types of agents engage in
distinct tasks and with diverse types of users. This is why the
\emph{context} layer of the KPM should always be accounted for,
especially when using the model for applied research.

The KPM also facilitates the exploration of risks associated with GSAs.
The unpredictability of black-box generative AI technologies can lead to
negative consequences, such as deception
\cite{ranischRapidIntegrationLLMs2025,singhInfluencePersuasiveTechniques2025}
or unsafe recommendations \cite{hundtLLMdrivenRobotsRisk2025}.
Awareness of these risks can help establish boundaries for the ethical
deployment of persuasive GSAs, ensuring their use aligns with users'
best interests and prevents exploitative practices. Moreover, insights
from the KPM might be used to increase persuasion literacy of human
users, for example by raising awareness of persuasive strategies
employed by GSAs \cite{markusImpactPersuasionLiteracy2025}. In addition,
technical guardrails need to be considered to protect users exposed to
potential irresponsible agent behavior and to allow human collaborators
to intervene when necessary \cite{rebedeaGuardrailsSecurityLLMs2025}.
Data protection is equally important, especially when GSAs rely on
sensitive user information
\cite{wangUniversityStudentsPrivacy2025,yaoSurveyLargeLanguage2024}.
Using locally run generative AI models may mitigate these risks
\cite{rebedeaGuardrailsSecurityLLMs2025}. Ensuring informed consent and
using secure, anonymized data storage protocols is also essential,
especially when user-specific knowledge informs the agent's behavior.
Adhering to societal norms and ethical standards is even more essential
when GSAs engage with vulnerable groups of users such as patients
\cite{elenduEthicalImplicationsAI2023}, children
\cite{gangulyConversationalAIAgents2026}, or the elderly
\cite{vonschallenKBDeldercare2026}.

While the KPM introduces new opportunities to studying persuasive
interactions with GSAs, two key challenges need to be addressed. First,
GSAs use generative AI systems capable of producing highly unrestricted,
context-sensitive outputs. Unlike agents with predefined actions, this
variability introduces difficulties in maintaining consistency and
comparability across participants. Hence, there is a need to measure the
consistency of a GSA's behavior within and between experimental
conditions. This can be done by using objective measures such as the
number and type of gestures used by the agent, length of agent messages,
or lexical similarities
\cite{irfanYouMeEthics2025,vonschallenKnowledgePowerImpact2026,vonschallenNeverSayNever2026}
as well as qualitative evaluations, for example to investigate the use
of persuasive strategies by the agent
\cite{vonschallenKnowledgePowerImpact2026,vonschallenNeverSayNever2026}.
A qualitative manipulation check may also prove important to investigate
whether differences in agent knowledge configurations actually lead to
observable differences in agentic behaviors
\cite{vonschallenKnowledgePowerImpact2026}. Second, the scope of the KPM
must be considered. The KPM was developed with a focus on understanding
persuasive GSAs, rather than human-driven persuasion, which has already
been extensively studied through established theories such as Cialdini's
seven principles \cite{cialdiniInfluenceSciencePractice2014}. Though
incorporating reciprocal interactions (i.e., the human user persuading
the agent) and multi-user dynamics (e.g., in debates) could broaden the
scope of the KPM, this expansion would also add complexity and
asymmetry, as human persuasion differs substantially from agentic
behavior generation. Nevertheless, future extensions of the KPM could
integrate these dimensions to extend the KPM to additional contexts.

To overcome these challenges, researchers should adopt best practices
tailored to the unique capabilities and risks of GSAs. Research
approaches, combining qualitative insights from conversation transcripts
with quantitative analyses are recommended for capturing the nuanced
relationships between variables of \emph{agent knowledge}, \emph{agent
persuasive behavior}, and \emph{user response}. Benchmarks might be
applied to assess the capabilities of generative AI models used in
interaction research in comparison to other models
\cite{changChatBenchStaticBenchmarks2025}. This could mitigate problems
related to comparability and generalizability. Additionally, upholding
open-science practices, realistic study designs, informed consent, and
robust data protection protocols will help ensure that research with
GSAs is ethically acceptable, replicable and contributes meaningfully to
both theoretical understanding and practical applications. When
upholding such methodological standards, the KPM can provide a
comprehensive framework for studying persuasive interactions with GSAs
across diverse contexts.

\section{Conclusion}\label{conclusion}

As technology continues to rapidly advance, a stronger understanding of
persuasive human--agent interactions will be vital for developing GSAs
that engage responsibly with human users. As autonomous agents become
increasingly persuasive \cite{holblingMetaanalysisPersuasivePower2025},
restricting their access to certain types of \emph{user-} and
\emph{context-knowledge} may help prevent malicious or manipulative
influence. At the same time, \emph{self-knowledge} instructions about
the agent's own limitations and barriers could reduce the risk of
providing arguments or suggestions outside the agent's own field of
competence. For example, equipping a GSA with \emph{self-knowledge}
about its own limitations may enable it to refrain from providing
medical or psychological advice and instead encourage users to seek
professional assistance, thereby reducing the risk of harmful persuasion
while promoting therapy compliance \cite{tannerSocialRobotsImprove2025}.
On the other hand, GSAs deployed for beneficial purposes -- such as
motivating users to adopt healthier lifestyles -- will require access to
knowledge that enables personalized, and context-sensitive persuasion
\cite{vonschallenKBDeldercare2026}. Studying
persuasive interactions with GSAs will be necessary to inform the
development of agents that motivate users for their own good, rather
than providing harmful or manipulative advice. The KPM provides a
comprehensive framework to study such interactions, ultimately promoting
responsible persuasion in alignment with human interest.

\bibliographystyle{plain} 
\bibliography{references}

\end{document}